\documentclass[]{cupconf}
\usepackage{graphicx}

\def\spose#1{\hbox to 0pt{#1\hss}}
\def\simlt{\mathrel{\spose{\lower 3pt\hbox{$\mathchar"218$}}
     \raise 2.0pt\hbox{$\mathchar"13C$}}}
\def\simgt{\mathrel{\spose{\lower 3pt\hbox{$\mathchar"218$}}
     \raise 2.0pt\hbox{$\mathchar"13E$}}}

\title[Chemical Evolution of the Galactic Bulge]
      {Chemical Evolution of the Galactic Bulge}
      
\author[B.K.~Gibson et al.]{\\ Brad K. Gibson$^1$ 
\\ Angela J. MacDonald$^1$
\\ Patricia S\'anchez-Bl\'azquez$^1$
\\ Leticia Carigi$^2$} 
\affiliation{$^1$ University of Central Lancashire, Centre for 
Astrophysics, Preston, PR1~2HE, U.K. \\ $^2$ Instituto de 
Astronom\'ia, U.N.A.M., M\'exico, D.F., M\'exico}

\begin{document}
\maketitle

\begin{abstract}
Adopting a single-zone framework, with accretion of primordial gas
on a free-fall timescale, the chemical evolution of the Galactic bulge 
is calculated, assuming (i) a corresponding rapid timescale for 
star formation, and (ii) an initial mass function biased towards massive 
stars.  We emphasise here the uncertainties associated with 
the underlying physics (specifically, stellar nucleosynthesis) and how
those uncertainties are manifest in the predicted abundance ratio 
patterns in the resulting present-day Galactic bulge stellar populations.
\end{abstract}

\firstsection
\section{Background}

In many respects, bulges of spiral galaxies are very similar to elliptical
galaxies.  Both adhere to many common scaling relations, including the 
Fundamental Plane, possess high stellar densities, little in the way
of gas and dust, and appear essentially old, with enhanced 
abundance ratios of $\alpha$-elements with respect to iron.  In our 
own Milky Way, the bulge accounts for $\sim$20\% of the Galaxy's
baryons - a factor of ten greater than the stellar halo.  Despite this
significance, relatively few detailed chemical evolution models of
the bulge exist,\footnote{In contrast to that of the halo, which 
despite its trace baryonic contribution to the Milky Way, has had
at least ten times the number of models published to 
explain its origin.} due in part to a dearth of 
high-resolution spectroscopic studies of its individual stars.
Having said that, in lieu of such data, notable exceptions have
appeared in the literature, drawing upon extant metallicity
distribution functions (MDFs) and inferred abundance ratios derived
from lower-resolution data.

K\"oppen \& Arimoto (1989,1990) assumed infall of primordial gas
on a freefall timescale (0.1~Gyr) and power-law initial mass functions
(IMFs) of slope $x$=1.05 (K\"oppen \& Arimoto 1989) and $x$=1.30
(K\"oppen \& Arimoto 1990), over the mass range 
0.05$<$$m$/M$_\odot$$<$60.  Rapid and efficient star formation
(10~Gyr$^{-1}$) was halted by a supernova-driven wind after
1~Gyr, the metal-enriched outflowing gas providing fuel to the 
Galactic disk for future star formation.  The K\"oppen
\& Arimoto models were successful in recovering the bulge MDF, 
present-day gas mass fraction, and enhanced [$\alpha$/Fe], despite
the (i) neglect of Type~Ia supernovae, and (ii) use of the 
instantaneous recycling approximation.  

Matteucci \& Brocato (1990) and, later, Matteucci et~al. (1999), 
relaxed these two limitations of K\"oppen \& Arimoto, also 
concluding
that flatter-than-Salpeter IMFs (1.1$\simlt$$x$$\simlt$1.3,
over the mass range 0.1$<$$m$/M$_\odot$$<$100) in
conjunction with (i) a Schmidt-like star formation law, (ii) rapid
infall of primordial gas on timescales of 0.01~Gyr (Matteucci
\& Brocato 1990) and 0.1~Gyr (Matteucci et~al. 1999), and (iii) 
efficient star formation (20~Gyr$^{-1}$).  Both models were 
successful in recovering the bulge MDF and enhanced [$\alpha$/Fe].

Samland et~al. (1997) suggested
that the bulge MDF is consistent with the use of a more traditional
Salpeter IMF ($x$=1.35, over the mass range 0.1$<$$m$/M$_\odot$$<$100) and
a more prolonged star formation
phase (with the bulge being 3-5~Gyr younger than the halo, a 
conclusion which is perhaps less secure), with "breathing" phases of
infall and outflow throughout the bulge's history.
Moll\'a et~al. (2000) also adopt the Salpeter IMF and assume
two infall phases ("bulge" and "core", with a longer infall timescale for
the dominant "bulge" phase of 0.7~Gyr).
As with Samland et~al.
(1997), infall and outflow leads to matter-exchange between halo, 
bulge, and core, and ultimately to a predicted bulge MDF which 
matches that observed.  

Each of the above models have their merits and detriments, but space
precludes a detailed intercomparison.  The prediction of $\alpha$-enhanced
abundance patterns {\it across the full range of bulge metallicities}
($-$1$\simlt$[Fe/H]$\simlt$$+$0.5) is somewhat unique to the 
"Matteucci" models, for obvious reasons (IMF + star formation efficiencies + 
timescales).

The recent appearance of spectacular high-resolution spectroscopic data 
for the bulge (e.g. Lecureur et~al. 2006, and references therein), 
makes it timely to revisit not only the traditional [$\alpha$/Fe]
patterns predicted by chemical evolution models, but also to begin to 
inspect individual $\alpha$-to-$\alpha$ element predictions, to seek
further insights into bulge formation (and, as we will suggest, 
stellar evolution).  Such a preliminary analysis was undertaken 
by Gibson (1995), but the data quality at the time made the
conclusions speculative, at best.
In this short contribution, we revisit the issue of bulge abundance
patterns, concentrating instead on a previously (somewhat) ignored
component of the models - specifically, the sensitivity to the adopted
Type~II supernovae yield compilation.

\section{Results \& Discussion}

Many of the models described in \S~1 adopt nucleosynthetic yields
from Woosley \& Weaver (1995), or one of its predecessors.
But, as discussed by Gibson et~al. (1997) in a different context
(clusters of galaxies), while stellar nucleosynthesis 
models may have identical global metal (i.e. $Z$) yields,
the relative distribution of elements and isotopes therein may be 
quite different (driven by differences in the
treatment of reaction rates, mass loss, convection, etc.).
Fig~1 (adapted from Gibson 1995) provides a graphic demonstration
for the important $\alpha$-element pair, oxygen and magnesium - a 
factor of 5-10 difference in O/Mg exists, for example, between
Woosley \& Weaver (1995) and Arnett (1991), at solar metallicity, 
in the mass range 15$\simlt$$m$/M$_\odot$$\simlt$25.

The shaded region highlights an interesting puzzle - as hinted at
already in the seminal work of McWilliam \& Rich (1994), and 
confirmed recently by Lecureur et~al. (2006),
O/Mg in the bulge appears to be a factor of two lower than
that of the Sun, over the metallicity range 
$-$0.5$\simlt$[Fe/H]$\simlt$$+$0.5 (i.e. a range spanning the bulk of
the stars in the bulge).  As discussed by Gibson (1995), such sub-solar
[O/Mg] values are essentially impossible to recover with {\it any}
chemical evolution model employing the Woosley \& Weaver (1995) yields,
as not a single model in the grid lies within the shaded region (and
thus, no IMF could be constructed \it a posteriori \rm which would
lead to a model matching the data); the situation does not appear 
particularly tenable with the Thielemann et~al. (1996) models either.
What {\it is} interesting from Fig~1 though is the location of 
the little-used Arnett (1991) yields in this particular plane - 
specifically, a natural byproduct of the models is 
the increased production of magnesium in the mass range
15$\simlt$$m$/M$_\odot$25, shifting those models into the 
abundance pattern regime populated by the stars of the Galactic
bulge (a not entirely surprising consequence, in light of Gibson 1997; 
Fig~3).

\begin{figure}
\begin{center}
\includegraphics[width=3.5in]{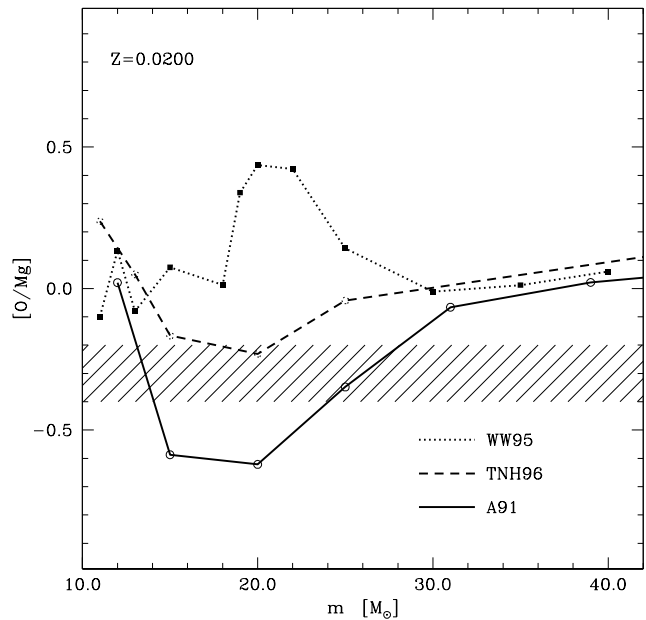}
\caption{Nucleosynthetic abundance ratio (oxygen-to-magnesium: [O/Mg])
patterns predicted by the 
solar-metallicity Type~II supernovae 
models of Woosley \& Weaver (1995: WW95), Thielemann
et~al. (1996: TNH96), and Arnett (1991: A91).  The shaded region
is representative of the range sub-solar [O/Mg] encountered in the 
Galactic bulge: $-$-0.4$\simlt$[O/Mg]$\simlt$$-$0.2 (McWilliam \&
Rich 1994; Lecureur et~al. 2006).}
\end{center}
\end{figure}

Using the Arnett (1991: A91), Woosley \& Weaver (1995: WW95), and 
Thielemann et~al. (1996: TNH96) Type~II supernova yields, we have 
constructed representative models of the Galactic bulge
using the one-zone infall precursor analog (Gibson \& Matteucci 1997)
to {\tt GEtool} (Fenner \& Gibson 2003).  Our fiducial bulge model is 
patterned after the ($x$=0.95, $k$=20~Gyr$^{-1}$, $\tau$=0.1~Gyr) model
of Matteucci et~al. (1999), the primary differences being (i) the
reduction of the IMF upper mass limit from 100~M$_\odot$ to 35~M$_\odot$,
and (ii) the inclusion of three different Type~II supernova yield
options.  Note, in keeping with our philosophy, we do 
{\it not} perform any {\it a posteriori} normalisation of our models.
As in K\"oppen \& Arimoto (1999,2000), we have stopped the simulations
at 1~Gyr, although this does not alter the general thrust of our
conclusions.

\begin{figure}
\begin{center}
\includegraphics[width=3.5in]{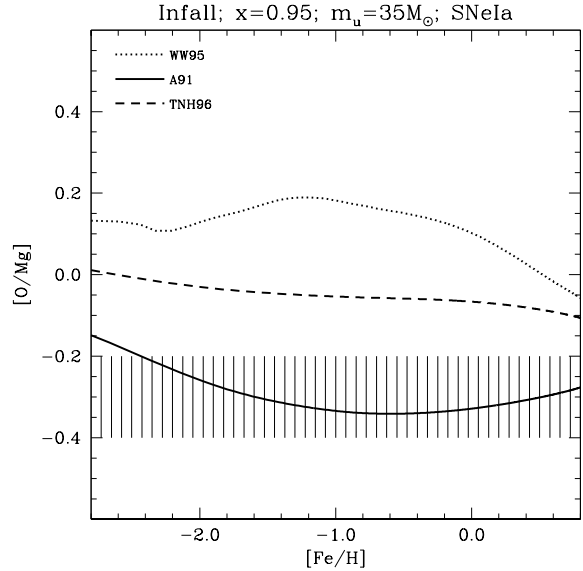}
\caption{Predicted chemical evolution trajectories in the [O/Mg]-[Fe/H]
plane, under the assumption of a massive-star biased initial mass function, 
rapid infall (on a free-fall timescale) of primordial gas, and the 
three Type~II supernovae yield compilations shown in Fig~1: 
WW95: Woosley \& Weaver (1995); TNH96: Thielmann et~al. (1996);
A91: Arnett (1991).  As in Fig~1, the shaded region
corresponds to the range sub-solar [O/Mg] encountered in the
Galactic bulge.}
\end{center}
\end{figure}

The predicted chemical evolution of this fiducial model, using the
three different yield sources, is shown in Fig~2.  It should come
as little surprise (in light of the discussion surrounding
Fig~1) to see that only the model incorporating Arnett's (1991) yields
successfully predict the bulk of the bulge stars to have 
$-$0.4$\simlt$[O/Mg]$\simlt$$-$0.2.  Conversely, 
Matteucci et~al. (1999; Fig~3) and Moll\'a et~al. (2000; Fig~3) predict 
[O/Mg]$\approx$$+$0.05$\pm$$+$0.05 for the stars of the Galactic bulge.
It should be stressed that this 
was a natural prediction of Arnett's yields, and required no
\it a posteriori \rm re-scaling of the magnesium abundances, as is
normally done when employing Woosley \& Weaver (1995).

Having said all this, it would be foolish to suggest that this is 
definitive proof in favour the Arnett (1991) compilation; all of
the caveats noted in Gibson et~al. (1997) regarding their input
physics remain valid today.  More importantly, what might ameliorate
one important abundance ratio problem in the bulge, may also lead to
irrepairable consequences for other patterns, or more likely, problems
for the solar neighbourhood (although a cynic could turn the problem
around and say that having to resort to a different IMF for the bulge,
as opposed to the solar neighbourhood, is not necessarily "better").
Indeed, we suspect that a detailed accounting of \it all \rm
relevant observables will suggest that the A91 yields are not a 
panacea for the chemical evolution of the bulge, but our goal here
was not to prove (or disprove) that statement, but simply to remind
the end-user of such yield tables 
that while WW95 is an extraordinarily beautiful suite of 
models, be cautious in assuming that their use constitutes the 
elimination of nucleosynthesis as a significant systemtic
uncertainty in models of galactic chemical evolution!

\begin{acknowledgements}
Simulations were performed at the University of Central 
Lancashire High Performance Computing Facility.
\end{acknowledgements}

\end{document}